\documentclass[preprint,showpacs,floats,letterpaper,floatfix,,groupedaddress,eqsecnum]{revtex4}
\usepackage{subfigure}

\usepackage{amssymb,amsmath}
\usepackage[dvips]{graphicx}

\usepackage{dcolumn,epsfig}

\begin{document}

\title{Stochastic Resonance in Periodic Potentials.}

\author{S. Saikia$^{1,2}$, A.M. Jayannavar$^3$ and Mangal C. Mahato$^1$}
\email{mangal@nehu.ac.in}
\affiliation{$^1$Department of Physics, North-Eastern Hill University, 
Shillong-793022, India\\
$^2$St. Anthony's College, Shillong-793003, India\\
$^3$Institute of Physics, Sachivalaya Marg, Bhubaneswar-751005, India}

\begin{abstract}
The phenomenon of stochastic resonance (SR) is known to occur mostly in 
bistable systems. However, the question of occurrence of SR in periodic 
potential systems is not conclusively resolved. Our present numerical work
shows that the periodic potential system indeed exhibits SR in the high 
frequency regime, where the linear response theory yields maximum frequency 
dependent mobility as a function of noise strength. The existence of two (and 
only two) distinct dynamical states of trajectories in this moderately 
feebly-damped periodically driven noisy periodic potential system plays an 
important role in the occurrence of SR.

\end{abstract}

\vspace{0.5cm}
\date{\today}

\maketitle

\section{Introduction}
Stochastic resonance (SR) was discovered theoretically about three decades 
ago\cite{Benzi}. Since then SR has been investigated with vigour and many 
notable reviews have appeared, for instance refer\cite{Gamma,Well}. Physical 
systems are always subject to internal or external (thermal or otherwise) 
fluctuations (noise). The optimal periodic response of a system to an external 
periodic drive as a function of noise strength is termed as stochastic 
resonance. It has been experimentally found to occur, to just mention a few,
in electronic circuits\cite{Fauve,Mantegna,Murali},
two-mode ring lasers\cite{Roy}, nanomechanical systems\cite{Mohanty}, neuronal
systems\cite{Douglass,Collins,Gluckman,Simonotto}, etc. Its main attraction 
lies in its practical utility of selecting and enhancing a signal of a 
particular frequency out of a host of signals by tuning the noise strength. 
Presumably, biological systems use SR to their advantage\cite{Moss}. It has 
the potential to be utilized to control kinetically the pathways of a 
biochemical reaction\cite{Ghosh}.

SR has been predicted and shown to occur mostly in bistable systems\cite{Gamma,
McN}. However, there has been some notable investigations of SR in monostable
and also periodic potential systems\cite{Stocks, Dykman}. In the monostable 
systems SR is shown to occur in the high frequency regime close to the natural 
frequency of oscillation at the bottom of the potential. However, the 
occurrence of SR in periodic potentials have not been conclusive\cite{Kim}.

Dykman and coworkers\cite{Dykman} use an interesting model in which a 
single-well Duffing oscillator with additive noise is driven at a frequency 
close to the natural frequency of the oscillator. The model is used under 
linear response theory formalism to study fluctuation phenomena associated 
with two coexisting periodic attractors. A weak Gaussian noise causes 
transition between these two attractors. The populations of these two 
attractors become equal at a particular noise strength where the response 
becomes maximum. This is considered as a genuine signature of a 
(nonconventional) stochastic resonance. The theoretical result was supported 
by an analog electronic circuit experiment. A similar result was obtained in 
an underdamped superconducting quantum interference device\cite{Kauf}.

The same resonance behaviour, as in Ref.\cite{Dykman}, in the frequency 
dependent mobility was obtained in a periodic potential using linear response 
theory by Kim and Sung\cite{Kim} in the high frequency range of the external 
periodic drive. However, these authors ascribe this resonant behaviour as 
simply a noise assisted standard dynamical resonance as the transitions 
involve only intra-well motion. Also, in the inter-well hopping (low) frequency 
regime the frequency dependent mobility shows monotonic behaviour as a 
function of noise strength thereby discounting the possibility of occurrence 
of SR in periodic structures. However, the authors show that SR can occur if 
the driven system has a tilted periodic potential so that the passages are
allowed only in one direction.

Moreover, recently, it was observed that in a bistable potential, $V(x)=
V_0e^{-ax^2}+b|x|^q/q$, the confinement parameter $q$ plays an important role
in deciding whether the system will show SR or not\cite{Hein}. For $q\geq2$ 
the system shows SR whereas for $q<2$ it does not. In addition, we find the 
input energy expended per period of the external field on the system by the 
field acts as a good quantifier of SR\cite{Iwa}. This input energy is 
ultimately dissipated into the thermal bath. This is naturally a measure of 
the hysteresis loop area in position($x$)-force($F$) space. Although the input 
energy and hysteresis loop area are exactly the same in magnitude the latter 
is an average quantity, whereas input energy has a well defined distribution. 
The input energy distribution provides useful information about stochastic 
resonance behaviour. In particular, the distribution shows a characteristic 
largest shoulder (bimodality) at stochastic resonance\cite{Saikia, Sahoo, Jop}.
On the other hand, hysteresis loops carry important information about phase
relationship between $x$ and $F$ which have also been of interest to SR\cite{
LGamma, Dykman1,Iwa}. 

In the present work, we explore the possibility of occurrence of SR in a 
periodic sinusoidal potential using input energy and hysteresis loop area as 
quantifiers. Moreover, one can take various values of the wavevector $k$ of
the potential analogously varying the effect of confinement parameter $q$
of the bistable potential discussed above. We, however, present results for
$k=1$ only.

We find that the periodic sinusoidal potential does not show SR when driven by 
a low frequency field corresponding to Kramers rates across the maxima of the
potential or when driven at a still lower frequency. The same conclusion have 
been arrived at in Ref.\cite{Kall} while studying the diffusion coefficient in
a periodic system. However, it should be noted that in Ref.\cite{Kall} the 
probability $P(\tau')$ of a particle, after going from one well to an adjacent 
one returns back to the same initial well in the subsequent time $\tau'$, shows 
periodic peaks. The strength of the first peak of $P(\tau')$ shows SR-like
behaviour. However, $P(\tau')$ has, by construction, the bearings of a bistable
system and not that of a periodic potential system.

We further find that in the high frequency range the input energy behaves 
exactly similarly as the response function in the works of Dykman and 
coworkers\cite{Dykman} and as the frequency dependent mobility does in the work
of Kim and Sung\cite{Kim}. In addition, our work shows that the input energy 
peaks as a function of noise strength. This is an indication of SR arising
due to a competition between two dynamical states of particle trajectories (to 
be elaborated in section III) as in the case of bistable systems. Though the 
trajectories are intra-well in nature close to SR the transition between these 
two states are also aided by inter-well passages of particles across the 
potential maxima. 

The two dynamical states of trajectories are distinctly identified by the 
phase difference $\phi$ between the periodic forcing $F=F_0\cos(\omega t)$ and 
the trajectory $x(t)=x_0\cos(\omega t+\phi)$; one having a fixed phase lag 
$\phi=\phi_1$ and the other $\phi=\phi_2$. Note that the system, at finite $T$,
being stochastic in nature $\phi_1$ and $\phi_2$ are average quantities. 
These individual phase lags $\phi_1$ and $\phi_2$ effectively do not vary with 
the noise strength. However, the relative cumulative length of these two 
dynamical states in a trajectory change with the noise strength. The average 
phase lag, therefore, changes with noise strength, similar
to what was predicted and observed in Ref.\cite{LGamma, Dykman1} in the case of 
bistable systems. Moreover, the distribution of input energy shows very 
similar behaviour across the input energy peak as in case of SR in a bistable 
system thus affirming the genuineness of SR in the present periodic potential 
system.

We consider two model systems for our study. In one case the medium is 
considered to having uniform friction, whereas in the other case the friction
is considered to be nonuniform. The nonuniform friction $\gamma(x)$ has an 
exact analogy in the Resistively Coupled Shunted Junction (RCSJ) model of 
Josephson junction. In the RCSJ model the '$\cos \phi$' term, arising out of
the interference between the quasiparticle tunneling and the Cooper pair
tunneling across the junction, is equivalent to the nonuniform part of 
$\gamma(x)$ in the present case. The systems with nonuniform friction, however,
are not very rare and are not without practical relevance\cite{Wanda}.  Since 
the potential is symmetric and periodic the homogeneous system does not show 
any average mobility. The nonuniform system, however, shows average current 
when driven by a sinusoidal forcing in the underdamped system\cite{Shantu}.

Interestingly, for the uniform system, $\phi_1\simeq -0.013\pi$ and 
$\phi_2\simeq -0.5\pi$ whereas in the nonuniform case, $\phi_1\simeq -0.025\pi$ 
and $\phi_2\simeq -0.85\pi$. In either case the trajectories have distributions
of these two phases depending on the noise strength. This is reflected in the
form of the $(x-F)$ hysteresis loops and hence in the average response 
amplitude and phase lag. In the homogeneous case the behaviour of the average 
phase lag only approximately conforms to the SR prediction of 
\cite{Dykman1,Iwa} and in the other case it follows closely the observations of 
\cite{LGamma} in bistable systems.

\section{The model}

In this work, we consider the underdamped motion of a particle in a 
periodic potential $V(x)=-V_0 \sin(kx)$ which is symmetric in space 
(about $kx=(2n+1)\pi/2$, $n=0, \pm1, \pm2, ...$).  The system is driven 
periodically by an external forcing $F(t)$=$F_0\cos(\omega t)$. We study two 
cases of the system -- when the friction coefficient $\gamma (x)$ is uniform 
(=$\gamma_0$) (system is homogeneous)  and when the friction coefficient is 
space dependent $\gamma (x)=\gamma_0(1-\lambda \sin(kx+\theta))$ (system is
inhomogeneous). In the latter case, the friction is periodic with the same 
periodicity as the potential but has a phase difference $\theta$  with it
($\theta \ne 0,\pi$). $\lambda$ ($0\leq\lambda<1$) determines the degree of 
inhomogeneity of the system ($\lambda=0$ corresponds to the homogeneous 
system).

\par A particle of mass $m$ moving in a periodic potential $V(x)=-V_0\sin(kx)$
in a medium with friction coefficient $\gamma (x)$ and subjected to an external 
periodic forcing $F(t)$ is considered to be described by the Langevin equation,
\begin{equation}
m\frac{d^{2}x}{dt^{2}}=-\gamma (x)\frac{dx}{dt}-\frac{\partial{V(x)}}{\partial
x}+F(t)+\sqrt{\gamma(x)T}\xi(t).
\end{equation}

\begin{equation}
m\frac{d^{2}x}{dt^{2}}=-\gamma_0\frac{dx}{dt}-\frac{\partial{V(x)}}{\partial
x}+F(t)+\sqrt{\gamma_0T}\xi(t).
\end{equation} 
Eqn. (2.1) is for the inhomogeneous system and Eqn. (2.2) for the homogeneous 
system. The temperature $T$ is in units of the Boltzmann constant $k_B$. 
The inherent random fluctuations in the system are represented by $\xi (t)$ 
which satisfy the statistics: $<\xi (t)>=0$, and 
$<\xi(t)\xi(t^{'})>=2\delta(t-t^{'})$. 
The equations are written in dimensionless units by setting $m=1$, $V_0=1$, 
$k=1$. The Langevin equation, with reduced variables denoted again now by the
same symbols, corresponding to Eqns. (2.1) and (2.2) are written as
\begin{equation}
\frac{d^{2}x}{dt^{2}}=-\gamma(x)\frac{dx}{dt}
+cos x +F(t)+\sqrt{\gamma(x) T}\xi(t).
\end{equation}

\begin{equation}
\frac{d^{2}x}{dt^{2}}=-\gamma_0\frac{dx}{dt}
+cos x +F(t)+\sqrt{\gamma_0 T}\xi(t).
\end{equation}
 
 The potential barrier between any two consecutive wells of $V(x)$ disappears
 at the critical field value $F_0=F_c=1$. The noise variable, in the same 
symbol $\xi$, satisfies exactly similar statistics as earlier.

\section{Numerical Results}
The trajectories $x(t)$ are obtained numerically\cite{Nume} by solving the 
Langevin equations (2.3) and (2.4) corresponding to the inhomogeneous and the 
homogeneous system, respectively, with the amplitude of the drive field, 
$F_0=0.2$ and $\omega=2\pi/\tau$, with $\tau=8$. The value of $\omega$ is 
close to the natural frequency of the potential. At high temperatures $T$ the 
average behaviour of these trajectories is the same independent of the initial 
conditions. However, at low temperatures, and especially in the limit of 
deterministic motion, and at such a low field amplitude as $F_0=0.2$, the 
trajectories are intra-well in nature. Yet, their behaviour is very sensitive 
to the initial conditions, $x(0)=x(t=0)$ and $v(0)=v(t=0)$\cite{Det}. In our 
numerical calculations we take $v(0)=0$ and $x(0)$ at $N$ equispaced intervals, 
$x_i$, $i=1,2,...,N$, between the two consecutive peaks, e.g. 
$[-\pi/2<x_i\leq 3\pi/2]$, of the periodic potential $V(x)$. In most of the 
cases we take $N=100$, but at some values of temperature we take $N=300$.

\begin{figure}[htp]
\centering
\includegraphics[width=7cm,height=10cm,angle=-90]{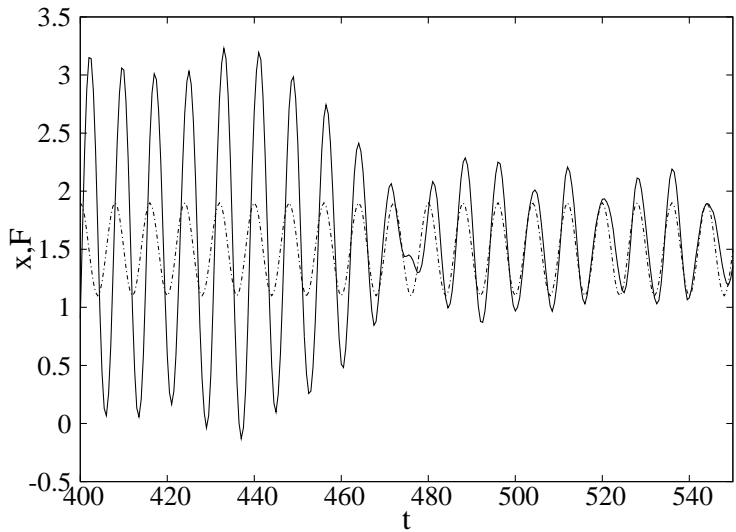}
\caption{Plot of $x(t)$ (solid line) and $F(t)$ (dot-dashed line) for $T=0.015$.
 The particle 
exhibits both kinds of trajectories with a transition from out-of-phase to 
in-phase, in this case, at around $t=460$; $\tau=8$, $F_0=0.2$, 
$\gamma_0=0.12$.}
\label{fig:edge}
\end{figure}

Depending on the range of $x(0)$ we get two distinctly different kinds of 
trajectories; one that lags behind the applied field by a small phase 
difference $\phi_1$ and the other by a large phase difference $\phi_2$. Only
for the sake of convenience, we call the former kind of trajectories as being
{\it{in-phase}} and the other as being {\it{out-of-phase}}. The out-of-phase
trajectories always have much higher amplitude than the in-phase trajectories, 
Fig.1. At the lowest temperature considered here there is no transition between
these dynamical states of trajectories. However, as the temperature (or the
noise strength) is increased transitions do take place, Fig.1. Yet, the 
trajectories are found basically only in these two states even at temperatures 
where inter-well transitions lead the trajectories out of the initial well.

Following stochastic enegetics formulation of Sekimoto\cite{Seki} the input
energy, or work done by the field on the system, $W$,  in a period $\tau$, is 
calculated as
\begin{equation}
W(t_0,t_0+\tau)=\int_{t_0}^{t_0+\tau}\frac{\partial U(x(t),t)}{\partial t}dt,
\end{equation}
where, the potential $U(x(t),t)=V(x)-xF(t)$, and $V(x)=-\sin(x)$, $F(t)=F_0
\cos(\omega t)$. The average input energy per period over an entire trajectory
$\overline{W}$, is
\begin{equation}
\overline{W}= \frac{1}{N_1}\sum_{n=0}^{n=N_1}W(n\tau,(n+1)\tau).
\end{equation}
Typically, the number of periods, $N_1$, taken in a trajectory ranges between
$10^5$ to $10^7$, as required. Finally, the average input energy per period 
$<\overline{W}>$ is calculated by averaging $\overline{W}$ over all the 
trajectories. From Eqn. (3.1) the distribution $P(W)$ is also calculated. As 
we shall see below $P(W)$ provides an important complementary criterion for 
stochastic resonance. 

Eqn.(3.1) can be further written as
\begin{equation}
W(t_0,t_0+\tau)=\int_{t_0}^{t_0+\tau}\frac{\partial U(x(t),t)}{\partial t}dt
=-\int_{F(t_0)}^{F(t_0+\tau)}xdF,
\end{equation}
which is the hysteresis loop area over a period. Since $x(t)$ is a stochastic 
variable it is unresonable to expect a sensible hysteresis loop over a period
of the field. However, when averaged over the entire duration of a trajectory
a well defined hysteresis loop $\overline{x}(F(t_i))$ and its area 
$\overline{A}$ is obtained:
\begin{equation}
\overline{x}(F(t_i))=\frac{1}{N_1}\sum_{n=0}^{n=N_1}x(F(n\tau+t_i)),
\end{equation}
for all $[0\leq t_i<\tau]$ and
\begin{equation}
\overline{A}=|\overline{W}|.
\end{equation}
The calculation of the hysteresis loops $\overline{x}(F(t_i))$, Eqn. (3.3), 
is correct at the lowest temperatures where the trajectories maintain the same 
phase $\phi$ throughout and also at higher temperatures where trajectories 
change phase between $\phi_1$ and $\phi_2$ during their journey. However, the
hysteresis loops will be different for different trajectories depending on the
cumulative duration of the dynamical states $\phi_{1,2}$ of the segments of 
the individual trajectories.
Therefore, an ensemble average $<\overline{x}(F(t_i))>$ is taken over all the
trajectories considered. The hysteresis loop $<\overline{x}(F(t_i))>$, with
area $<\overline{A}>$, reflects the average response $x(t)$ of the system to 
the applied field $F(t)$. In the linear response regime one would expect 
$x(t)=x_0\cos(\omega t+\overline{\phi})$ for $F(t)=F_0\cos(\omega t)$, 
$\omega=2\pi/\tau$, where $\overline{\phi}$ is the average phase lag to be 
observed experimentally\cite{LGamma}. Notice that $\overline{\phi}$ will, in 
general, be different from the fixed phases $\phi_1$ and $\phi_2$ 
characterising the individual dynamical phases of the trajectories.

In the following, we present the results of our numerical calculations and
analyse how the particle trajectories in the sinusoidal potential and hence
the input energy, and the hysteresis loops are affected by noise strength. We
examine the occurrence of stochastic resonance in the periodic potential,
referring to the standard SR criteria in bistable systems.
\subsection{Homogeneous Systems}

\begin{figure}[htp]
  \centering
  \subfigure[]{\includegraphics[width=6cm,height=9cm,angle=-90]{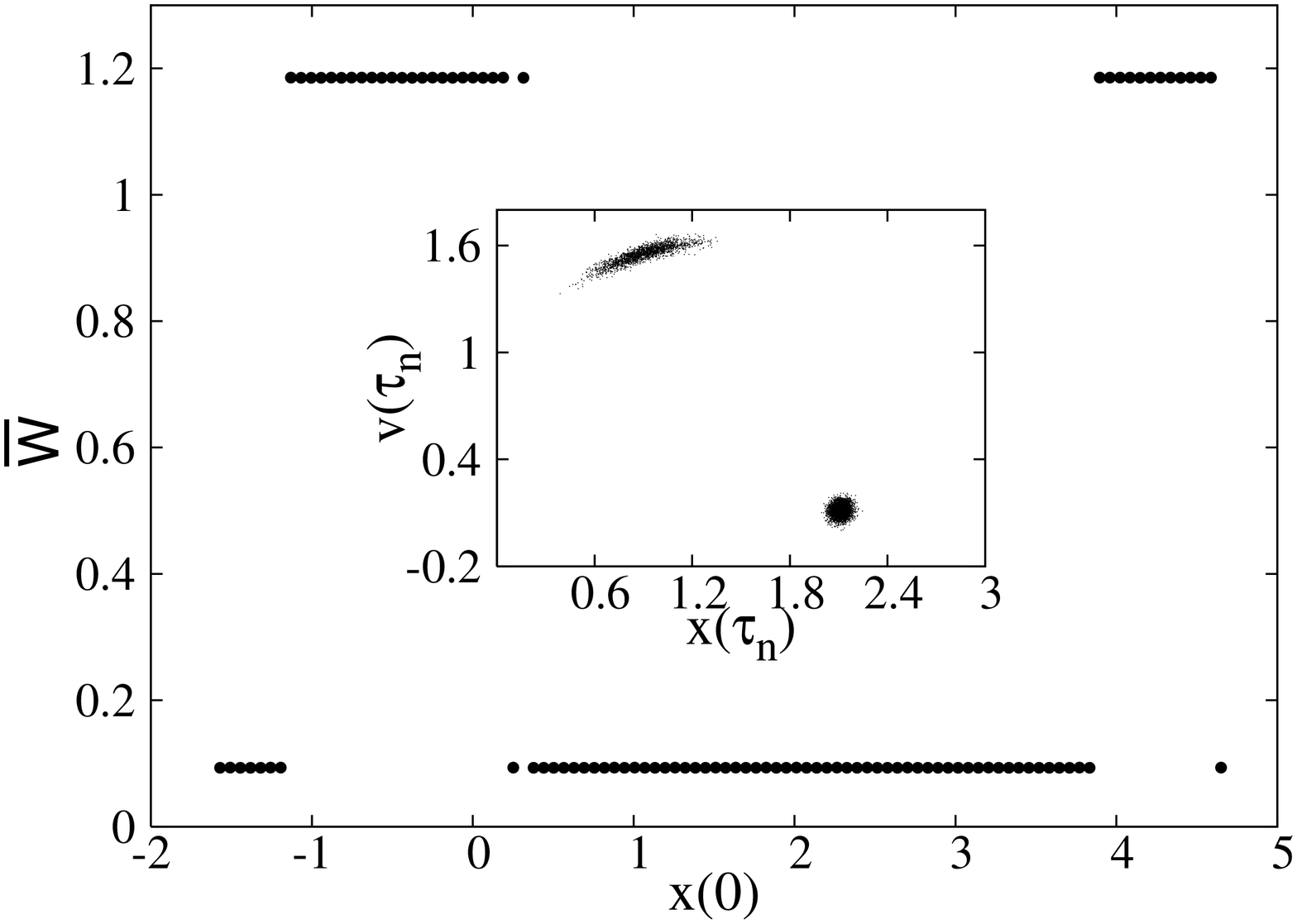}}
\hspace{0.4cm}
\subfigure[]{\label{fig:edge-c}\includegraphics[width=6cm,height=9cm,angle=-90]
{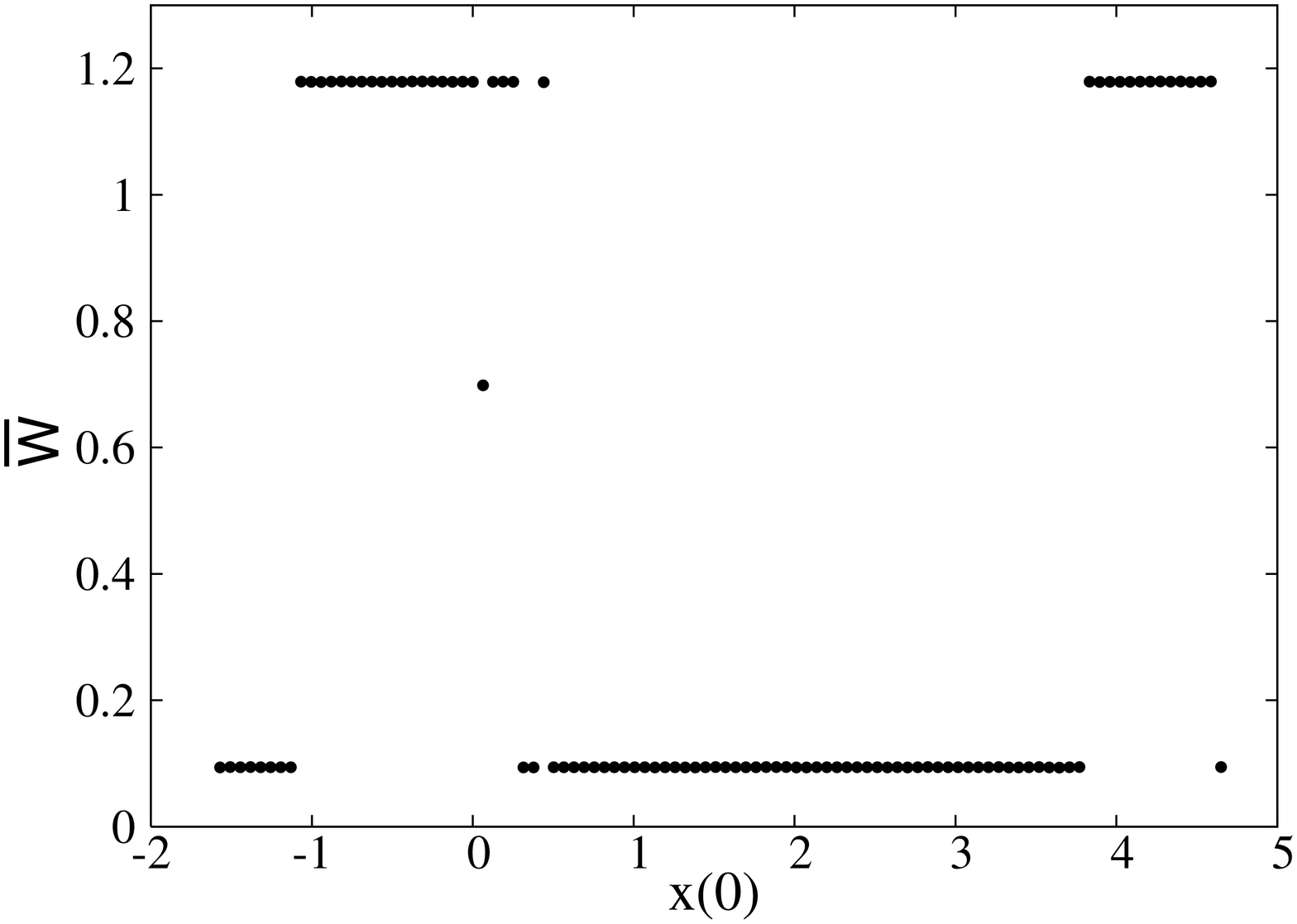}}
\hspace{0.4cm}
\subfigure[]{\includegraphics[width=6cm,height=9cm,angle=-90]{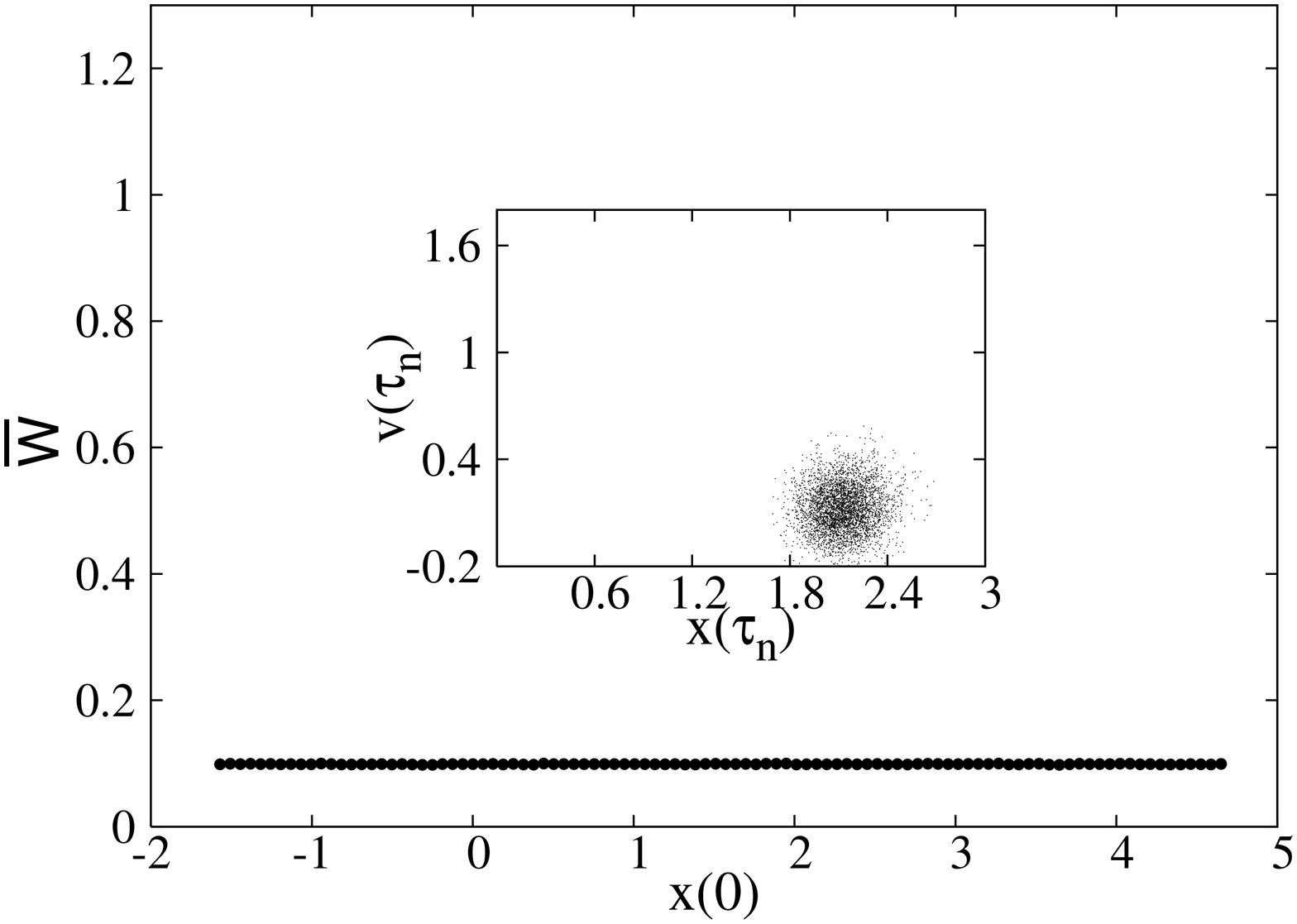}}

\caption{Plot of $\overline{W}$ with $x(0)$ for $T=0.001$ (a), $T=0.003$ (b) 
and $T=0.016$ (c), $\tau=8$, $F_0=0.2$, $\gamma_0=0.12$. The  insets in Figs. (a)
and (c) show the stroboscopic (Poincar\'{e} plots in the $(x-v)$ plane at times 
$\tau_n=n\tau, n=0, \pm1, \pm2$, ... .}
\label{fig:edge}
\end{figure}

\subsubsection{The intra/inter-well transitions and input energy}
Fig.2a shows the plot of input energy $\overline{W}$ averaged over a 
trajectory with initial position x(0) at the lowest temperature $T=0.001$ for
the homogeneous system ($\gamma(x)=\gamma_0$). $\overline{W}$ are confined to 
two narrow bands around 0.094 and 1.179 corresponding to $x(0)$ values that lie
in contiguous regions in the range $[-\pi/2,3\pi/2]$ and no points in between.
The input energies $\overline{W}=0.094$ correspond to in-phase trajectories 
with phase lag $\phi_1\simeq-0.013\pi$ and those with $\overline{W}=1.179$ 
correspond to out-of-phase trajectories with phase lag $\phi_2\simeq-0.5\pi$.
The trajectories continue to be in the same state of phase lag throughout
their course of journey. At this low temperature the trajectories are quite 
stable. The inset of Fig.2a shows the $(x-v)$ plane Poincar\'{e} 
(stroboscopic) plots revealing the two attractors corresponding to the two
dynamical states of the trajectories. The (stochastic) spread in the plots is 
due to the finite temperature of the system.

It may be recalled that the results are obtained by taking $v(0)=0$. However, 
the two dynamical states of trajectory are genuinely stable as revealed, in
Fig.3, by the basins of attraction of the two attractors at $T=0.001$. In the 
figure the 
in-phase and the out-of-phase states are indicated respectively by inp($l$) 
and outp($m$) for $l,m=0, \pm1, \pm2$, etc. The indices $l,m$ (within the 
braces) specify the well number where the particle settles down as it begins
from the initial well ($l,m=0$). Even though the attractors correspond to
wells $l,m\ne 0$, the phase and amplitude of the trajectories remain exactly
the same as when in the initial well ($l,m=0$). Thus, the differently labeled
trajectories are physically the same two and the only two: the in-phase and
the out-of-phase with fixed $\phi_1$ and $\phi_2$, respectively.  

Fig.2b shows the same plot as Fig.2a but at $T=0.003$. The figure clearly 
shows that a 
few points have deserted the upper band of $\overline{W}$ and all of them but
one have already joined and the lone one on its way to join the lower band of 
$\overline{W}$. This shows that all of these few out-of-phase trajectories 
have made a transition to the in-phase state (as shown in Fig.1) quite early 
during the $10^5$ periods of the trajectory and the lone one somewhere in the 
middle of all these periods. It is to be noted that no transition has taken 
place from the in-phase state to the out-of-phase state. This shows that the 
out-of-phase state is less stable than the in-phase state and are separated by 
an energy barrier of about $0.003$ (in units of $V_0$, the amplitude of the 
sinusoidal potential) from the out-of-phase side. This trend continues till 
$T=0.016$ where $\overline{W}$ consists of only one band, that is the lower 
band, Fig.2c. The stroboscopic plot in the $(x-v)$ space, inset of Fig.2c, 
shows the lone attractor at $T=0.016$. The other attractor has now ceased to 
exist. From $T=0.003$ to $T=0.016$ the ensemble averaged input energy 
$<\overline{W}>$ decreases very rapidly attaining the lowest value at 
$T=0.016$.

\begin{figure}[htp]
\centering
\includegraphics[width=8cm,height=12cm,angle=-90]{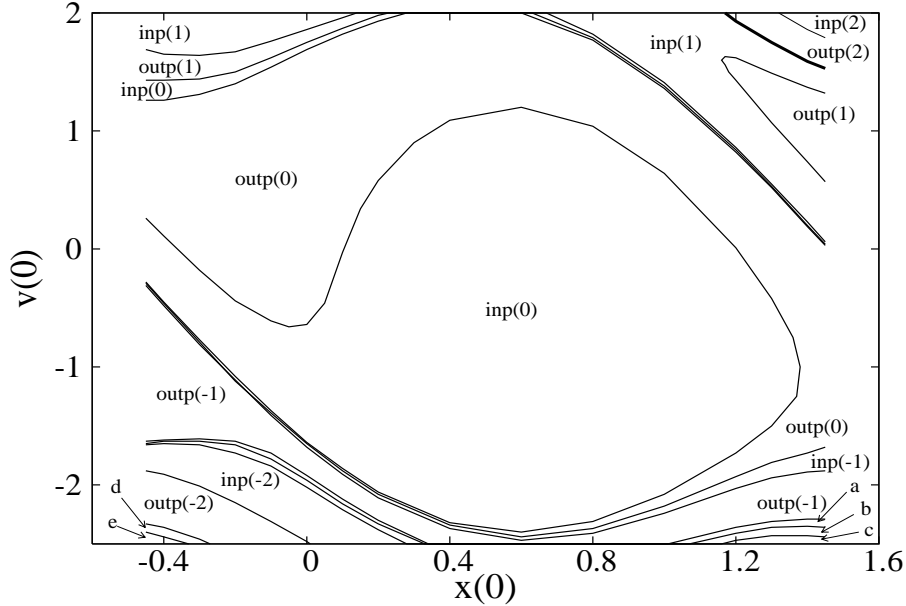}
\caption{The Basins of attraction of the in-phase inp and out-of-phase
outp attractors at temperature $T=0.001$ are shown. The bracketed numbers
on inp and outp indicate the well number of the periodic potential where
the trajectory settles down in the in-phase or out-of-phase states. The region
marked 'a' correspond to inp(-1), 'b' correspond to outp(-2), 'c' to the continuation
of inp(-2) indicated on the left side, 'd' to inp(-3), and the region 'e'
corresponds to outp(-3). The boundaries correspond to the end of the phase 
indicated above the respective lines. The thick line separating outp(2) and
 inp(1) indicates that
some other phases too chip in between. The phase inp(1) at the top right corner is
just the continuation of the phase inp(1) on the left top corner of the figure.
 For this figure, $\tau=8$, $F_0=0.2$, $\gamma_0=0.12$.}
\label{fig:edge}
\end{figure}

\begin{figure}[htp]
\centering
\includegraphics[width=7cm,height=10cm,angle=-90]{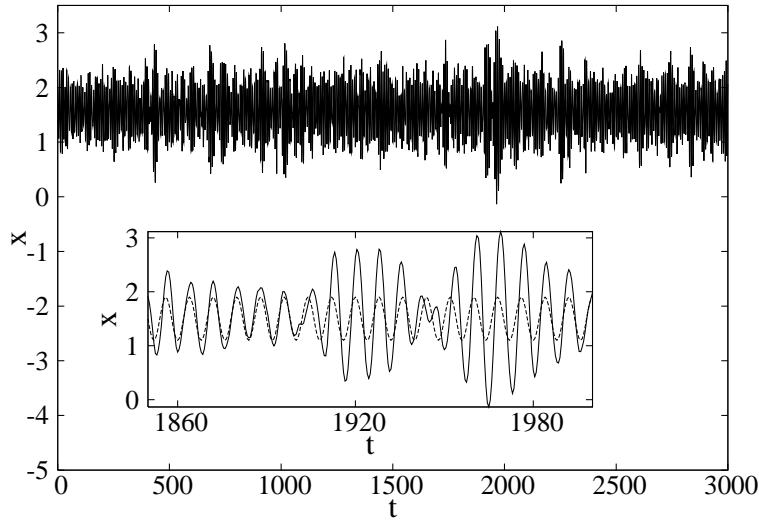}
\caption{Plot of the particle trajectories $x(t)$  for $T=0.04$.
Transistions are seen between the in-phase to the out-of-phase states.
Inset is a magnified figure. $F(t)$ (dashed line) is also included for comparison;
$x0=2.0$, $\tau=8$, $F_0=0.2$, $\gamma_0=0.12$.}
\label{fig:edge}
\end{figure}

\begin{figure}[htp]
\centering
\includegraphics[width=6cm,height=9.5cm,angle=-90]{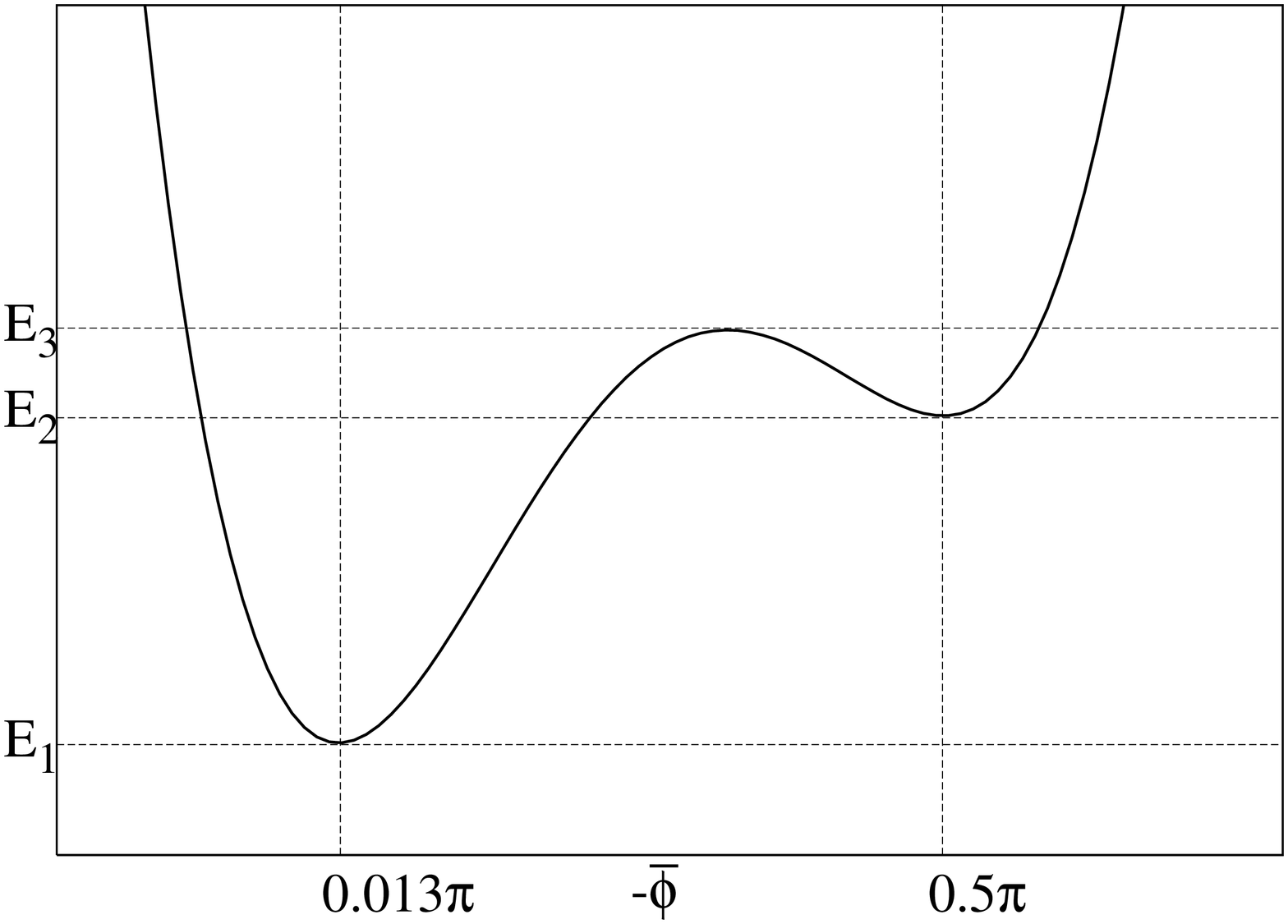}
\caption{ A schematic depiction of the two dynamical energy states of the
particle trajectories. In the figure $E3-E1\simeq0.016$ and $E3-E2\simeq0.003$.}
\label{fig:edge}
\end{figure}

For $T>0.016$, transitions start taking place from the in-phase to the 
out-of-phase state too. Consequently, $<\overline{W}>$ begins rising for 
$T>0.016$. Of course, the out-of-phase state lives for a very short duration 
before a larger temperature is reached. This is demonstrated in Fig.4, for 
$T=0.04$. Therefore, there is an energy barrier of roughly about 0.016 from 
the in-phase state side to the out-of-phase side. One can thus roughly 
pictorise the two states as shown in Fig.5 with the bottom of the wells at 
$-\phi_1=0.013\pi$ and $-\phi_2==.5\pi$.

\begin{figure}[htp]
\centering
\includegraphics[width=7cm,height=10cm,angle=-90]{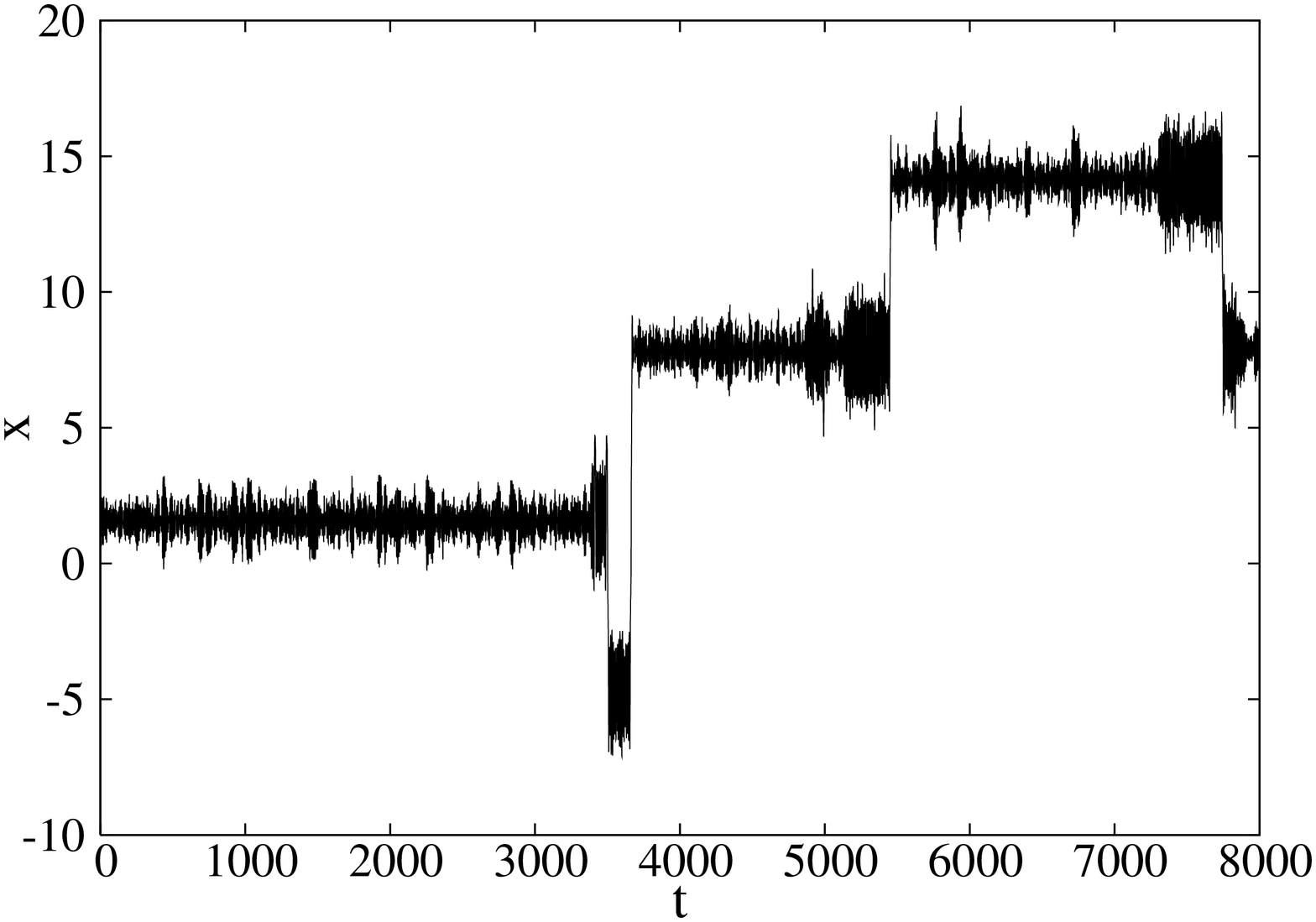}
\caption{Plot of $x(t)$ for $T=0.08$. The figure shows inter-well transistions 
as also numerous transistions between the in-phase (lower amplitude) and 
out-of-phase (higher amplitude) states. At around $t=3700$ and $t=5400$, the 
inter-well jump also leads to a transistion from the out-of-phase to the 
in-phase state; $x(0)=2.0$, $\tau=8$, $F_0=0.2$, $\gamma_0=0.12$.}
\label{fig:edge}
\end{figure}

As the temperature is increased further the relative population of out-of-phase
state keeps increasing with increase of temperature. The transitions in both 
the directions maintain a constant ratio at any given temperature. Moreover,
by the temperature $T\sim 0.08$ the inter-well transitions have already set in
and have become numerous. These inter-well transitions also help intra-well
transitions, Fig.6. At around $T=0.2$, the relative population of both the 
states become almost equal and the input energy $<\overline{W}>$ peaks. 
$T=0.2$, thus, falls in the region of kinetic phase transition\cite{Dykman2} 
between the two dynamical states.

\begin{figure}[htp]
  \centering
  \subfigure[]{\includegraphics[width=6cm,height=7cm,angle=-90]{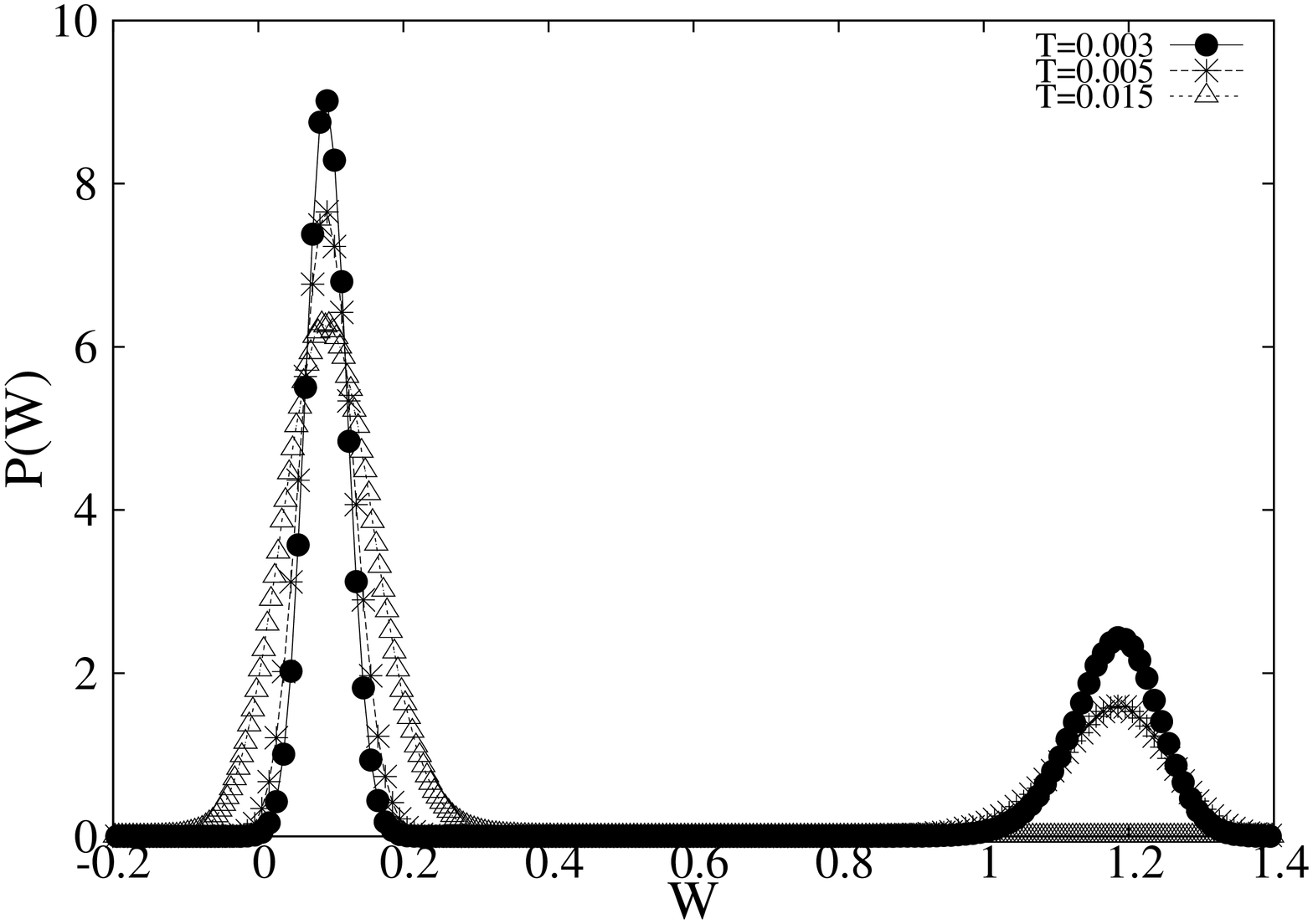}}
\hspace{0.4cm}
\subfigure[]{\label{fig:edge-c}\includegraphics[width=6cm,height=7cm,angle=-90]
{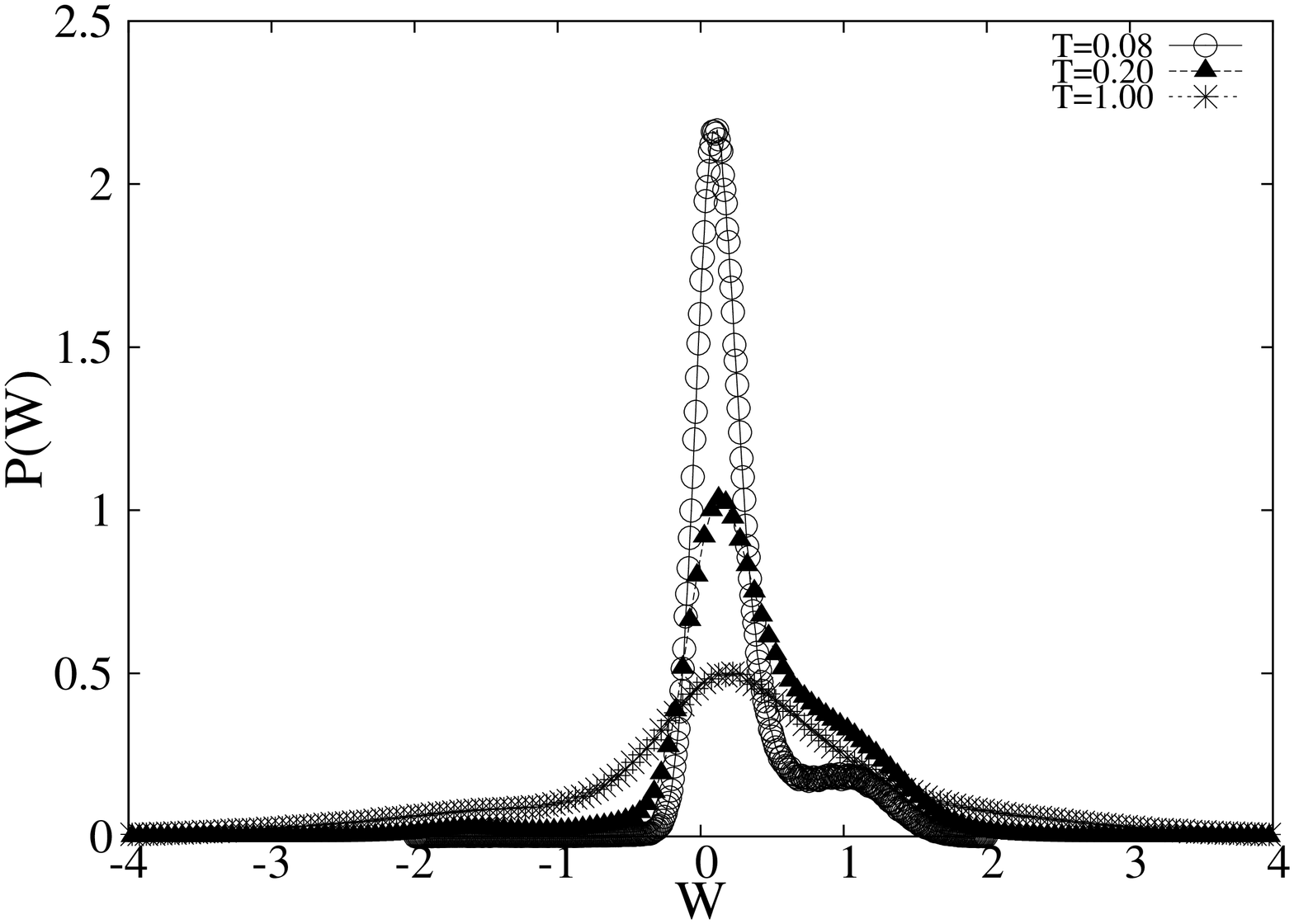}}
\caption{Plot of $P(W)$ for different values of $T$ (Fig. a for $T<0.016$ and
Fig. b for $T>0.016$) for the homogeneous system.
For large $T$, e.g., at $T=1.0$, the distribution has a single peak structure;
$\tau=8$, $F_0=0.2$, $\gamma_0=0.12$.}
\label{fig:edge}
\end{figure}

Beyond $T=0.2$, the intra-well transitions become more frequent and so do the
inter-well transitions. The transition points or the attempted transition 
points lead to lower amplitude motion. This results in gradual lowering of 
$<\overline{W}>$ with temperature. At much higher temperatures $T\gg0.2$ the
(intra-well as well as inter-well) transitions become so numerous that the
phases effectively lose their distinct identity. This gets reflected in the 
input energy distribution $P(W)$ as a single peak (Fig.7, $T=1.0$).

In Fig.7 the probability distribution of $W$ is drawn for various temperatures.
At the lowest temperature $T=0.003$ we see two distinct peaks of $P(W)$: The
low $W$ peak corresponding predominantly to the in-phase state of trajectory
and the other to the out-of-phase state. As the temperature is gradually 
increased the large $W$ peak shrinks and at $T=0.016$ this peak disappears 
completely. However, as the temperature is increased further the out-of-phase
peak reappears and begin to swell to maximise at $T\sim0.2$. At $T=0.2$ the
broad shoulder of $P(W)$ characterises the peak in $<\overline{W}>$. At the
largest temperature shown $T=1.0$ both the peaks merge into a broad single 
peak. Though $<\overline{W}>$ always remains $>0$ thus never violating the
second law of thermodynamics, interestingly, $P(W)$ is not confined to $W>0$
but a significant part of it lies at $W<0$. 

\begin{figure}[htp]
\centering
\includegraphics[width=7cm,height=10cm,angle=-90]{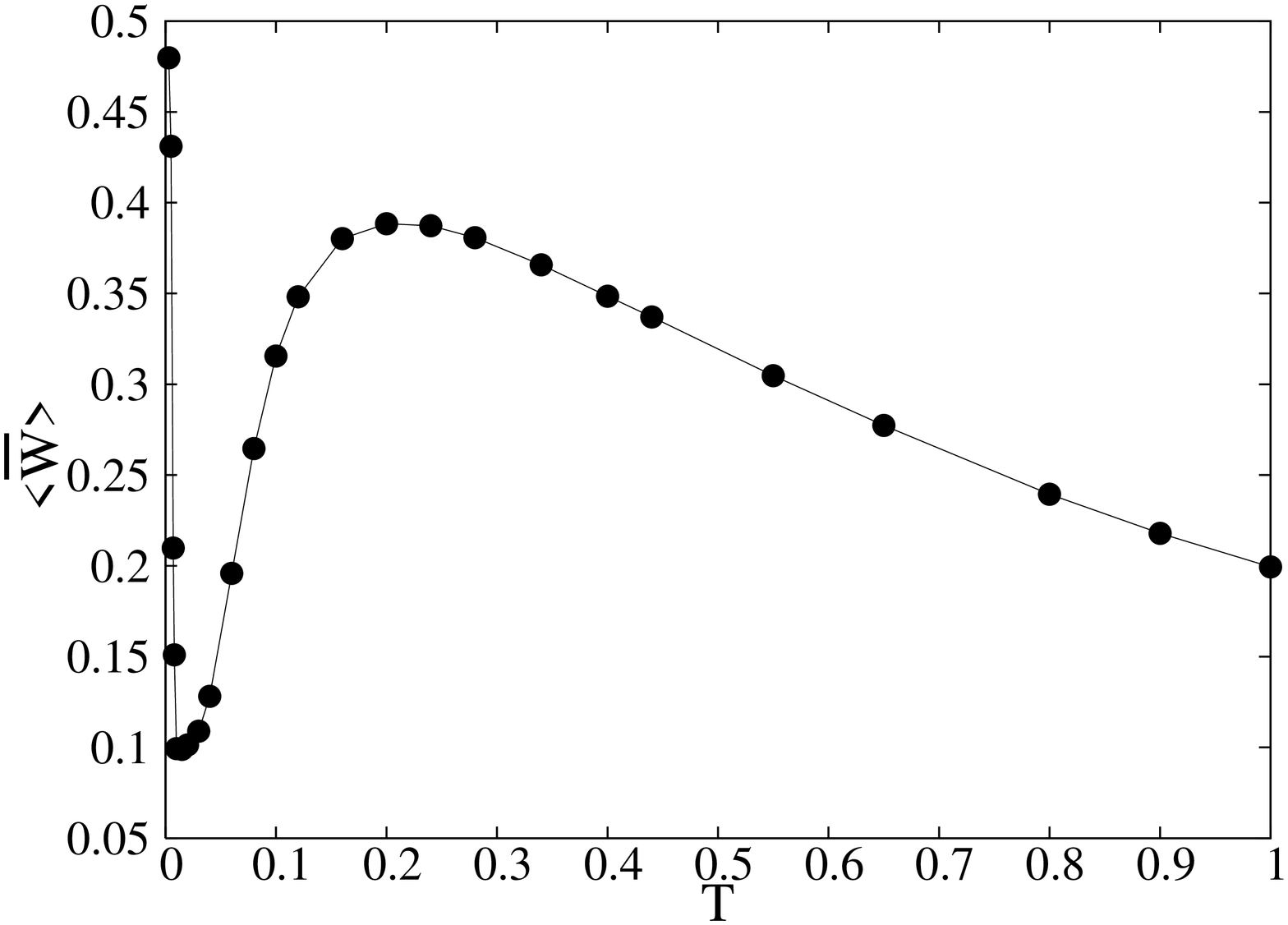}
\caption{Plot of $\langle \overline{W} \rangle$ as a function of $T$ for the 
homogeneous system;
 $\tau=8$, $F_0=0.2$, $\gamma_0=0.12$.}
\label{fig:edge}
\end{figure}

Fig.8 shows the variation of $<\overline{W}>$ as a function of temperature $T$.
$<\overline{W}>$ peaks at a temperature around $T=0.2$. This is a clear 
signature of stochastic resonance even if the criterion for SR in a bistable 
system is adhered to as a benchmark\cite{Hein, Saikia,Sahoo}. This is 
supported by the behaviour of $P(W)$ across $T=0.2$. $P(W)$ shows a prominent
shoulder\cite{Saikia,Sahoo}, characteristic of SR, around $T=0.2$, Fig.7. In 
the following we examine the behaviour of phase lag of the response to the 
periodic field.

\begin{figure}[htp]
\centering
\includegraphics[width=18cm,height=9cm,angle=-90]{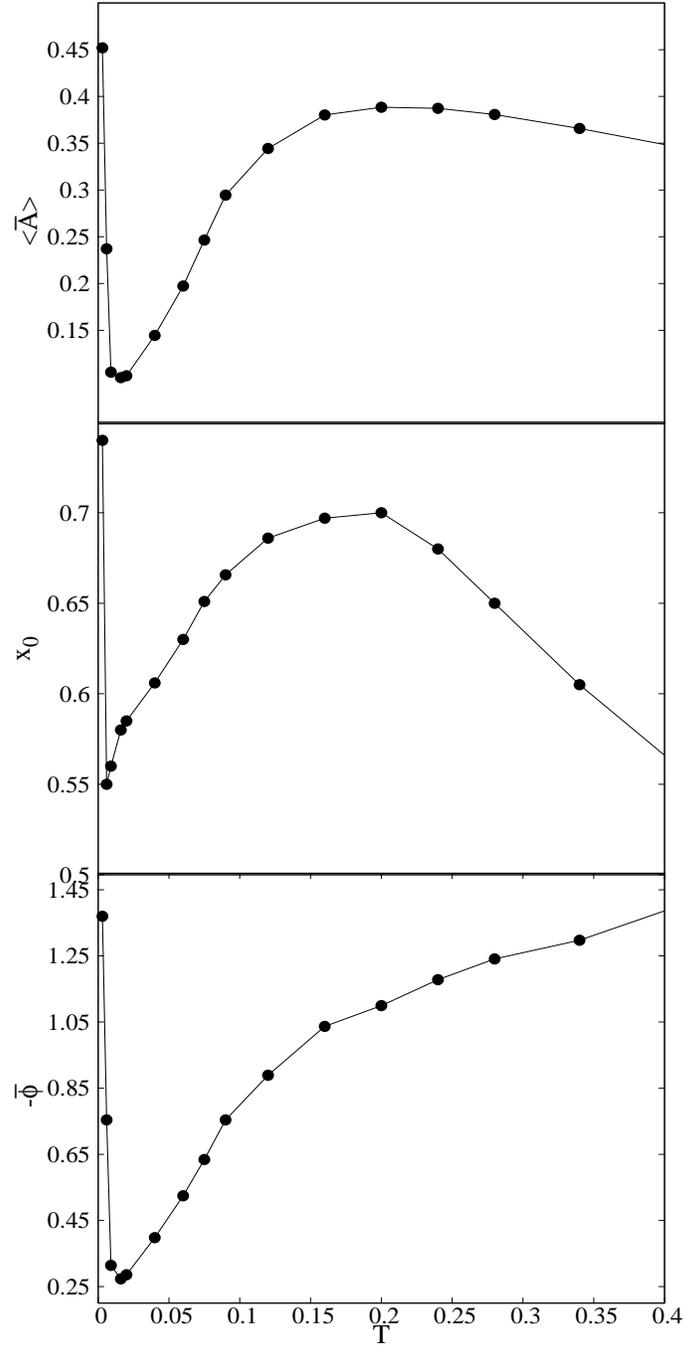}\hspace{0.4cm}
\caption{Plot of $\langle \overline{A} \rangle$ (top), $x_0$ (middle) 
and $-\overline{\phi}$ (bottom) 
with $T$ for the homogeneous system; $\tau=8$, $F_0=0.2$, $\gamma_0=0.12$. 
Beyond $T=0.4$, the curves have a monotonic behaviour and are not shown. 
The top figure is an exact reproduction of Fig.8.}
\label{fig:edge}
\end{figure}

\subsubsection{The Hysteresis loop: Area and the phase}
From Eqns. (3.3), (3.4) and (3.5) the hysteresis loop area $<\overline{A}>$ 
being equal in magnitude to $<\overline{W}>$, $<\overline{A}>$ does not provide 
any additional information than $<\overline{W}>$. However, the hysteresis loop
itself does give an important insight. The amplitude $F_0=0.2$ of the external
applied field $F(t)=F_0\cos(\omega t)$ being small, one expects the response, 
namely the position $x(t)$, to have a linear variation with $F(t)$:
$x(t)=x_0\cos(\omega t+\phi)$. Hence, the hysteresis loop 
$<\overline{x}(F(t_i))>$ will closely be an ellipse. The average phase 
difference $\overline{\phi}$ is measured from the resulting ellipse, Eqn. 
(3.4). Unlike $\phi_1\simeq -0.013\pi$, and $\phi_2\simeq -0.5\pi$, which 
remain more or less constant with $T$, the average phase difference 
$\overline{\phi}$ varies with $T$, as shown in Fig.9. In Fig.9, the average 
amplitude $x_0$ and the average area $<\overline{A}>$ are also plotted for 
comparison. 

At the lowest temperature $T=0.001$, $\overline{\phi}$ is close to 
$\phi_2\simeq -0.5\pi$. At this temperature, $x_0$ as well as $<\overline{A}>$ 
are also at their respective maxima. As $T$ is increased $-\overline{\phi}$ as 
well as $<\overline{A}>$ decrease sharply, and attain a minimum at around 
$T=0.016$. The minimum $\overline{\phi}$ is close to $\phi_1\simeq -0.013\pi$. 
This is because at $T=0.016$ all the trajectories are in the dynamical state 
of phase $\phi_1$. Thereafter for $T>0.016$, $-\overline{\phi}$ increases 
monotonically. The inflection point of $-\overline{\phi}(T)$ occurs at 
$T\simeq 0.09$, where $-\overline{\phi}\simeq \pi /4$ as observed in 
Ref\cite{LGamma}. However, $<\overline{A}>$ becomes maximum only at a much 
higher temperature $T\simeq 0.2$, where $-\overline{\phi}\simeq 0.35\pi$. 
Therefore, $<\overline{A}>$ here does not exactly satisfy the additional 
approximate SR criterion on $\overline{\phi}$ suggested in Ref\cite{LGamma}. 
However, in the inhomogeneous system, the SR criterion on $\overline{\phi}$ 
suggested using linear response theory\cite{Dykman1} appears to be respected.

\subsection{Inhomogeneous Systems}
In this case the particle experiences a nonuniform friction as it moves in the
medium. As stated earlier the potential is considered periodic: 
$V(x)=V_0\sin(kx)$. The friction coefficient $\gamma(x)$ is taken as 
$\gamma(x)=\gamma_0(1-\lambda\sin(kx+\theta))$ instead of the constant friction
coefficient $\gamma(x)=\gamma_0$ as in the homogeneous case. We take fixed
values $\lambda=0.9$ and phase difference $\theta=0.35$ throughout. 
$\theta(\neq 0,\pi)$ provides the necessary asymmetry in the system to yield
an average particle current, the ratchet current\cite{Det}, even when driven 
by a zero time-average external forcing $F(t)=F_0\cos(\omega t)$. For $k=1$, 
$\gamma_0=0.12$, $F_0=0.7$, and $T=0.4$ the system shows a current maximum at
$\omega=2\pi/140$. Keeping $\omega=2\pi/\tau, \tau=140$ and other parameters 
fixed the current maximizes at $T\simeq 0.25$. However, we do not see any 
maximum in the periodic response, such as the hysteresis loop area, to the 
external periodic field corresponding to these parameters. The system does not 
show SR at this low frequency of drive. 

Here, the particle position $x(t)$ does roughly follow the periodic field with 
an irregular but large amplitude. There is also a small amplitude high 
frequency component superimposed on the low frequency response to the field. 
The frequency of the superimposed component is close to the natural frequency
of the potential. In the following we consider the external field $F(t)$ with
$F_0=0.2$ and $\tau=8$, exactly  as in the homogeneous case, but keeping in 
mind that in the inhomogeneous case the system is no longer symmetric. 
Therefore, even at this high frequency the system shows an average ratchet 
current but the current itself does not show any peaking behaviour with $T$.

\subsubsection{The intra/inter-well transitions and input energy}
As in the homogeneous case, the system shows two distict dynamical states of 
trajectories. One with phase difference $\phi_1\simeq-0.025\pi$ and the other
with $\phi_2\simeq -0.85\pi$. We again call the trajectories with phase 
difference $\phi_1$ as being in-phase and the other as out-of-phase with the 
external field. The latter having much higher response amplitude $x_0$ than 
the former. The frictional nonuniformity, surprisingly, leads to out-of-phase 
trajectory amplitude about three times larger than in the homogeneous case and 
therefore larger average energy $<\overline{W}>$. As a consequence, the results
are qualitatively different from what was observed in the case of uniform 
friction system.

\begin{figure}[htp]
  \centering
  \subfigure[]{\includegraphics[width=6cm,height=7cm,angle=-90]{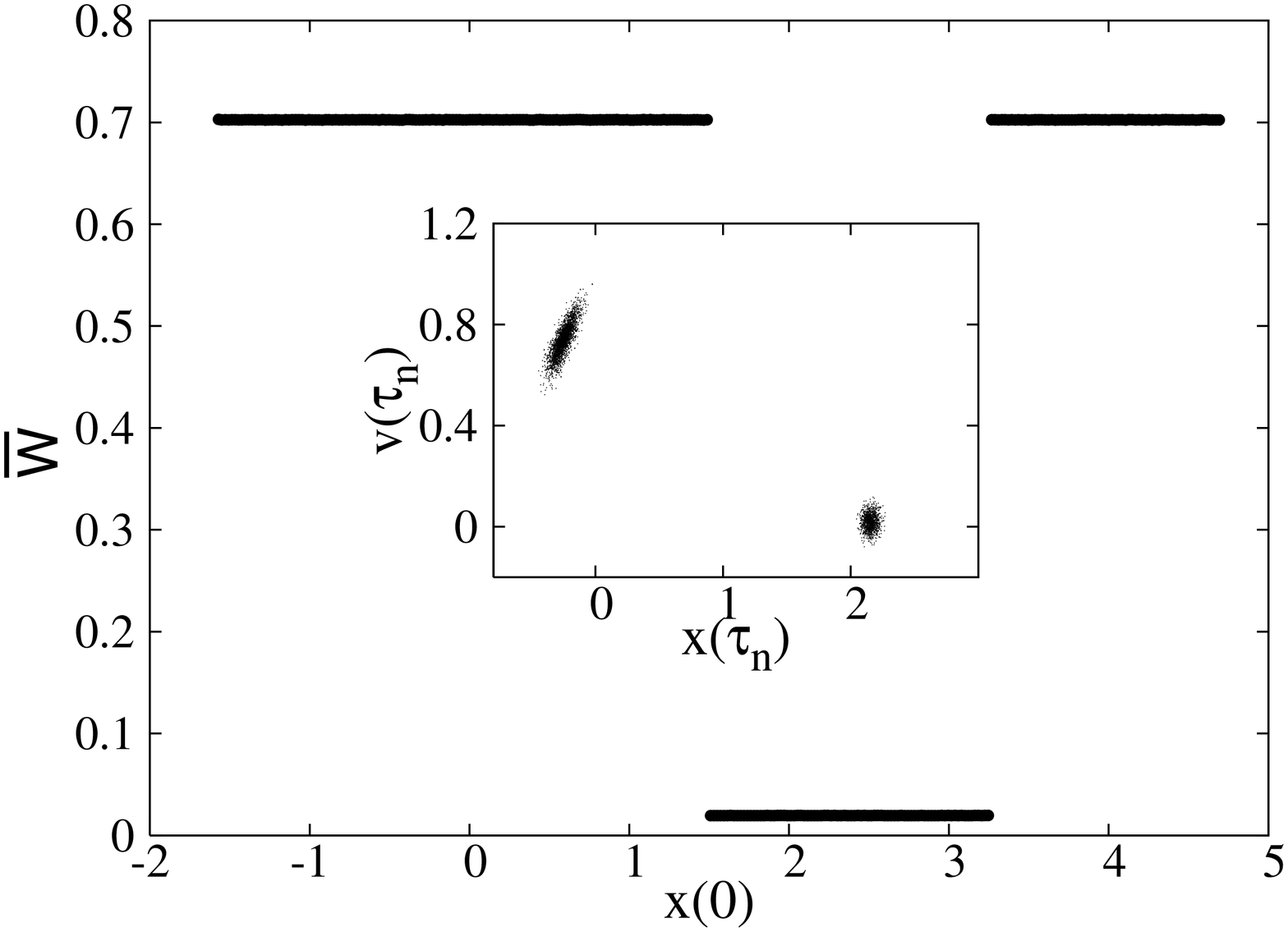}}
\hspace{0.4cm}
\subfigure[]{\label{fig:edge-c}\includegraphics[width=6cm,height=7cm,angle=-90]
{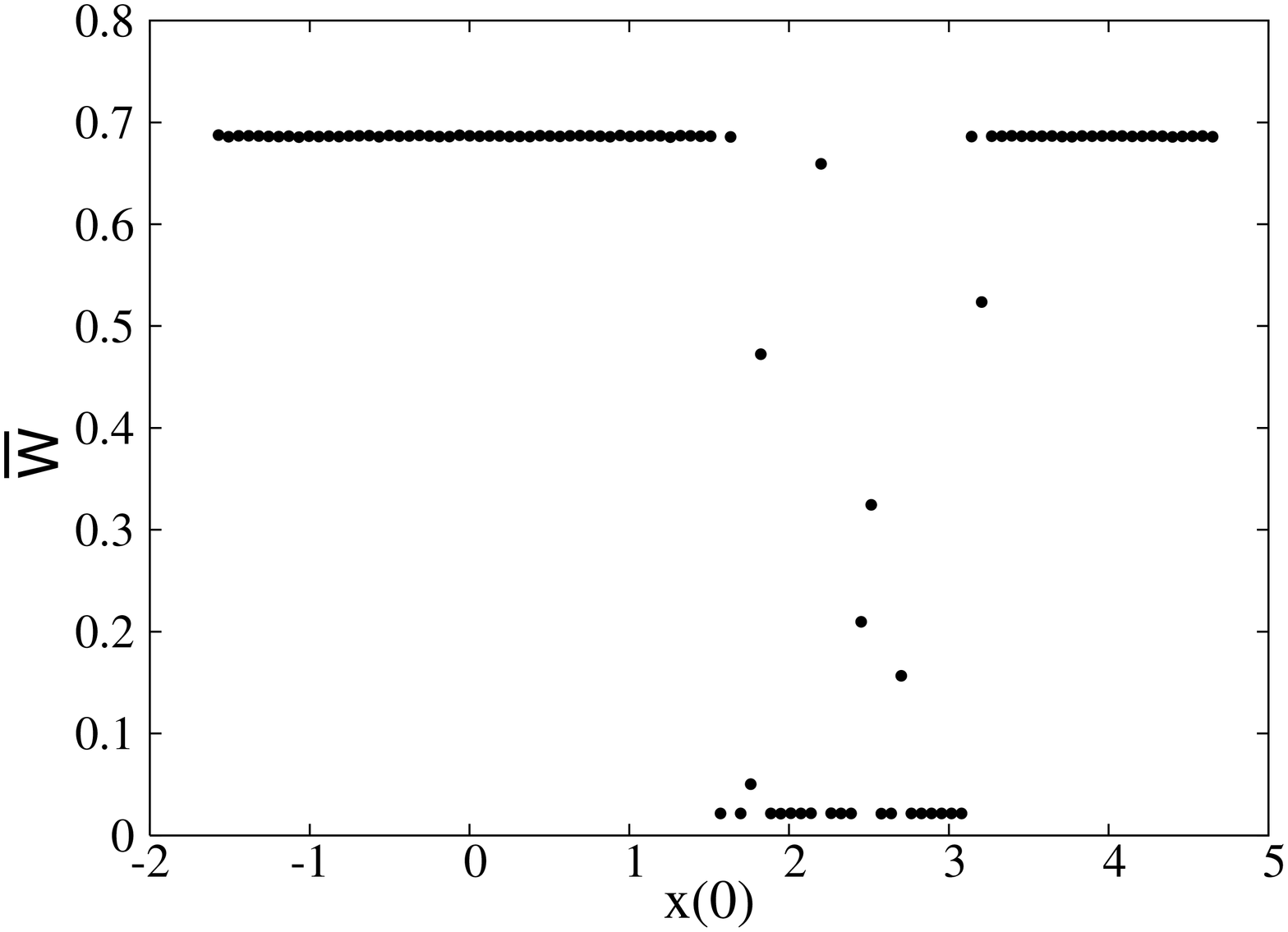}}
\hspace{0.4cm}
\subfigure[]{\includegraphics[width=6cm,height=7cm,angle=-90]{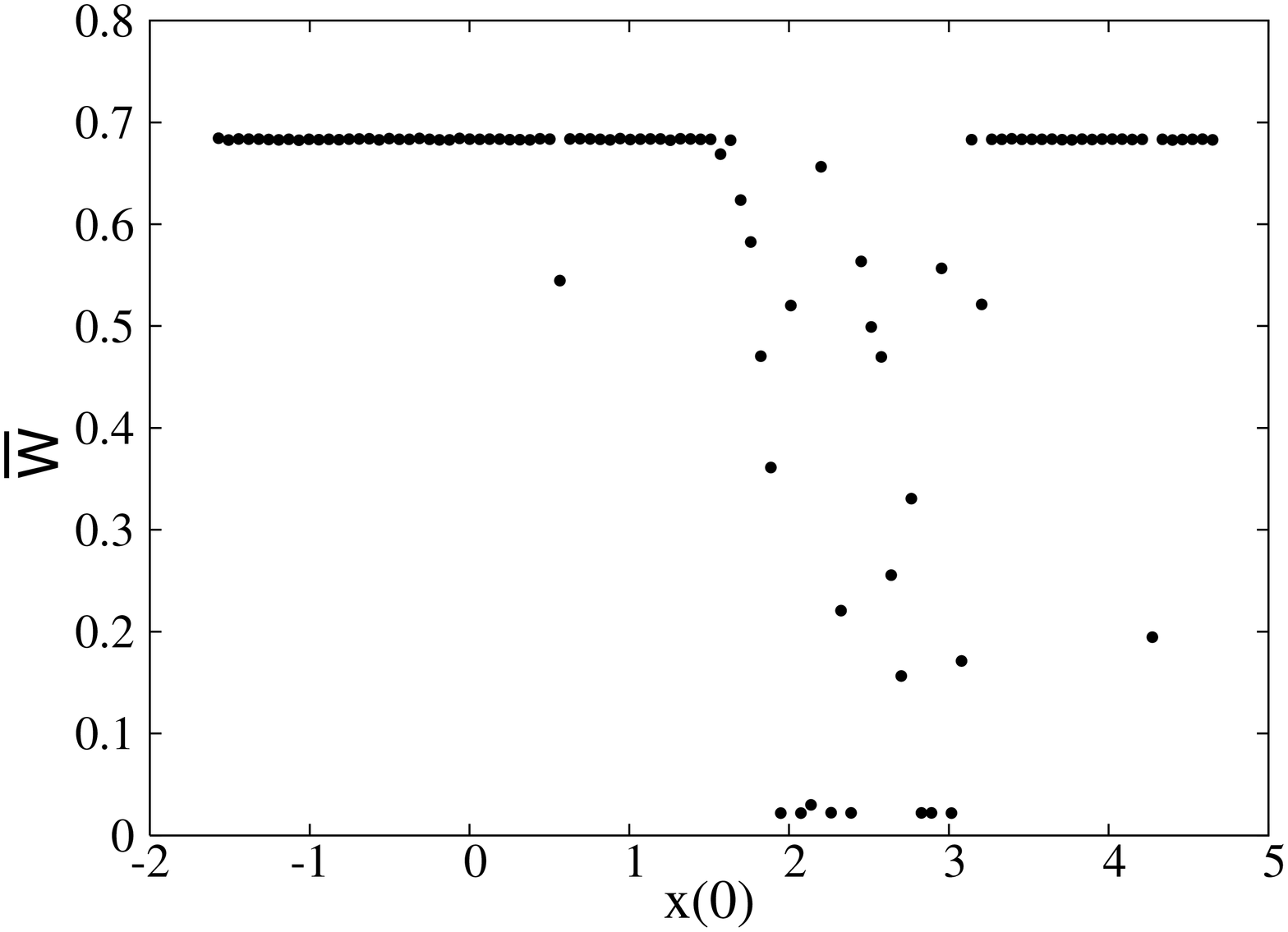}}
\hspace{0.4cm}
\subfigure[]{\includegraphics[width=6cm,height=7cm,angle=-90]{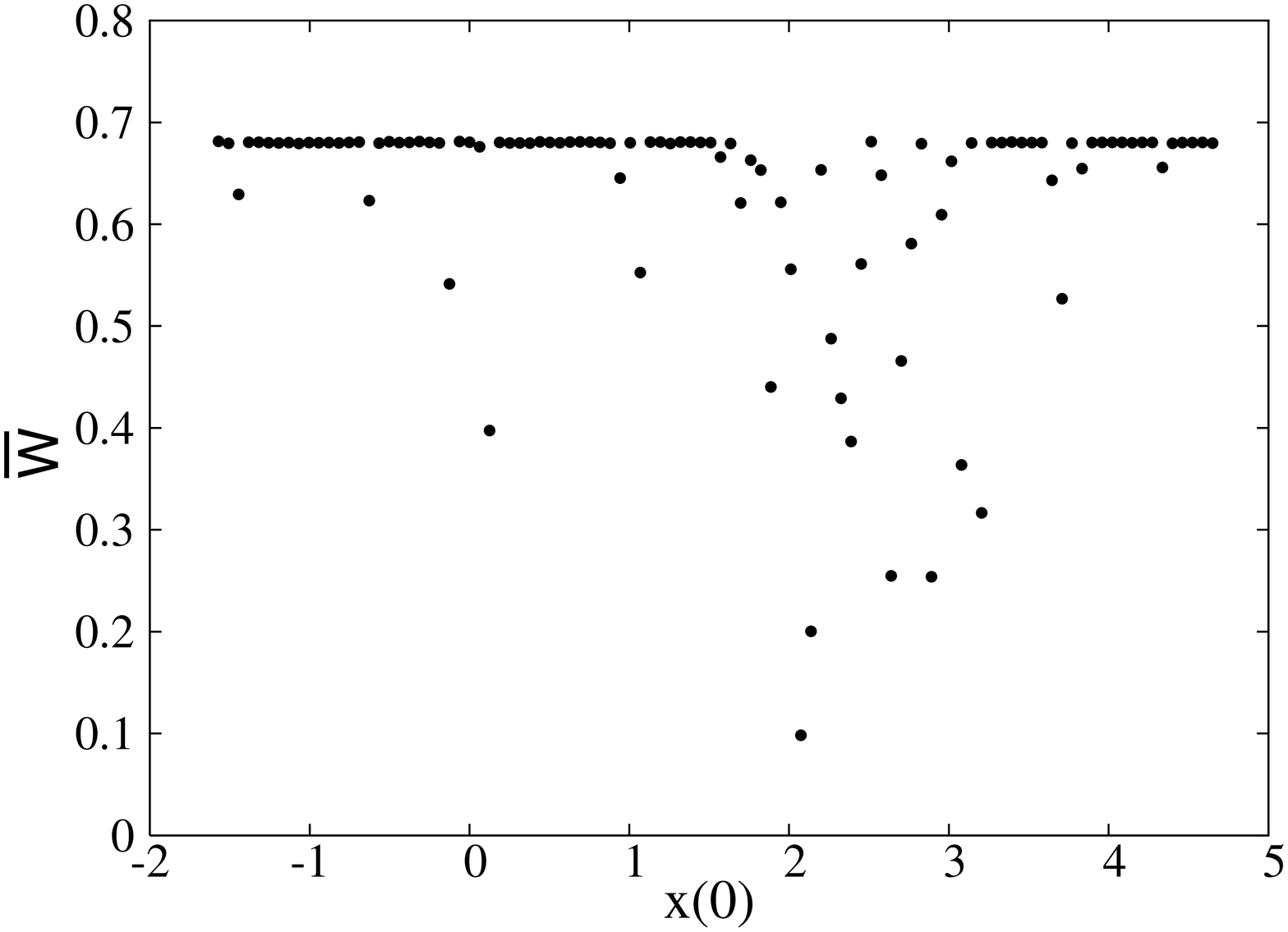}}
\caption{Plot of $\overline{W}$ with $x(0)$ for $T=0.001$ (a), $T=0.012$ (b), 
$T=0.014$ (d) and $T=0.016$ (d), $\tau=8$, $F_0=0.2$, $\gamma_0=0.12$ for the 
inhomogeneous system. The
insert of Fig.a shows the two attractors in the $(x-v)$ plane stroboscopic
plots at times $n\tau$.}
\label{fig:edge}
\end{figure}

\begin{figure}[htp]
\centering
\includegraphics[width=15cm,height=9cm,angle=-90]{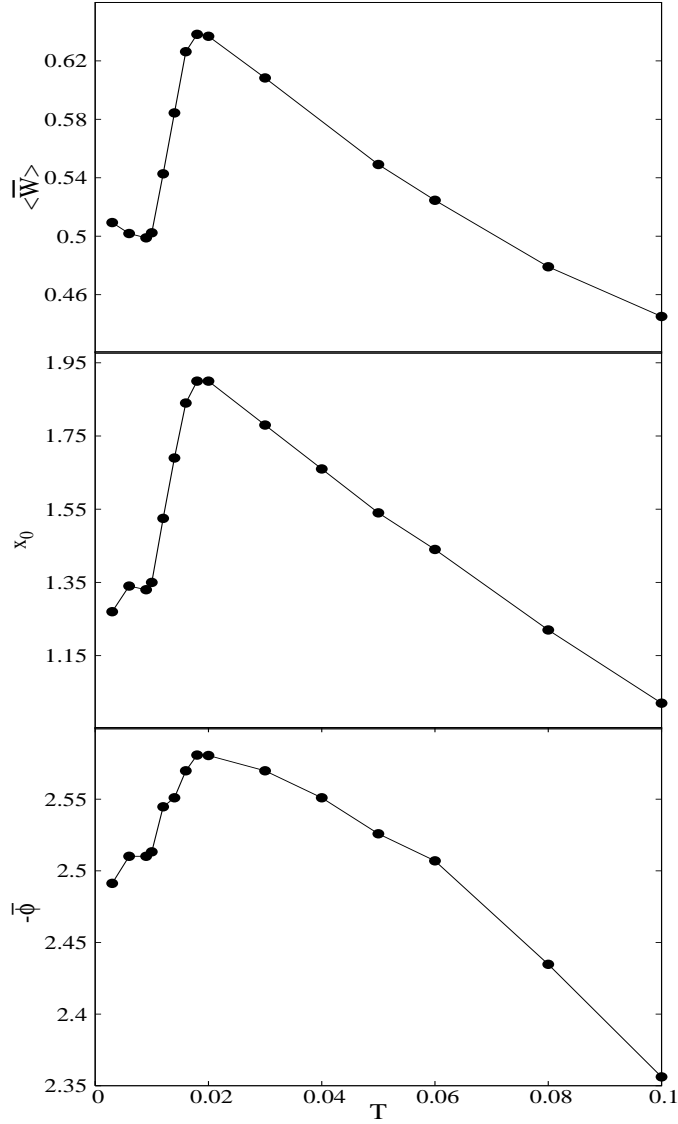}\hspace{0.4cm}
\caption{Plot of $<\overline{W}>$ (top), the amplitude $x_0$ (middle) and 
the phase difference $-\bar \phi$ (bottom) with $T$ for the inhomogeneous system;
 $\tau=8$, $F_0=0.2$, $\gamma_0=0.12$.}
\label{fig:edge}
\end{figure}

At the lowest temperature $T=0.001$ the trajectories are bunched into two
contiguous groups and continue to be so for all time without ever jumping
across the groups, Fig.10a. The inset of Fig.10a shows, in the $(x-v)$ plane
stroboscopic plots, the two attractors corresponding to the two dynamical 
states of particle trajectories. The situation remains the same till $T=0.01$.
However, the average energy $<\overline{W}>$ decreases with $T$ due to a
slight decrease in the out-of-phase trajectory amplitude. Contrast this with 
the sharp decline in energy in the homogeneous case due to transition from
out-of-phase states to the in-phase states. At $T$ around 0.011 transitions 
begin to take place from the in-phase states to the out-of-phase states, 
Fig.10b; the out-of-phase state is more stable. The $<\overline{W}>$ rises 
sharply thereafter. However, by $T=0.014$ transitions are also observed from 
out-of-phase states to the in-phase states, Fig.10c. Therefore, the potential 
barrier from out-of-phase side to the in-phase side is about 0.011, whereas in 
the reverse direction the barrier height is about 0.014. Both the states are 
{\it{almost}} equally stable. By the temperature $T=0.016$ even inter-well 
transitions are observed, Fig.10d.

\begin{figure}[htp]
  \centering
  \subfigure[]{\includegraphics[width=6cm,height=7cm,angle=-90]{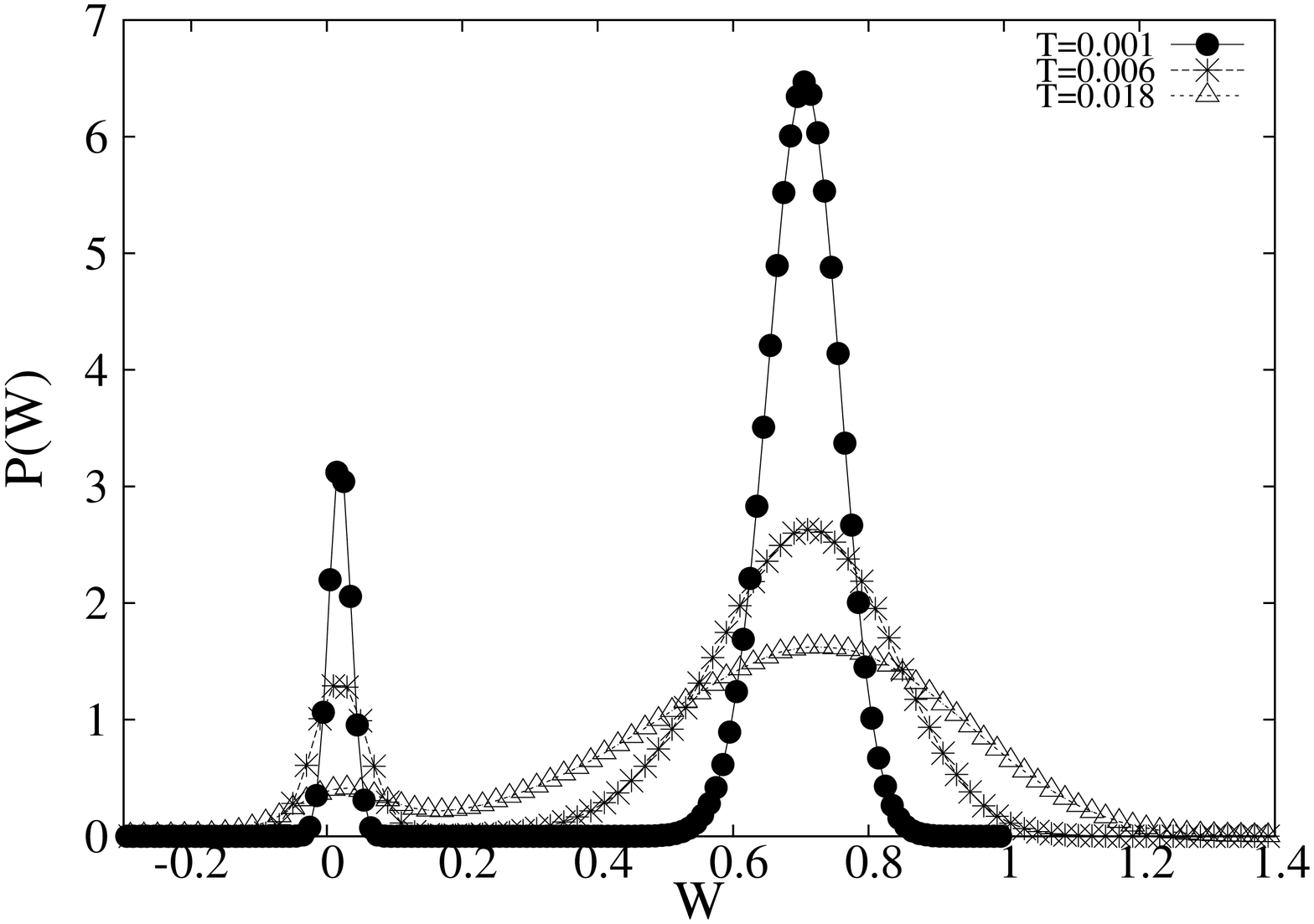}}
\hspace{0.4cm}
\subfigure[]{\label{fig:edge-c}\includegraphics[width=6cm,height=7cm,angle=-90]
{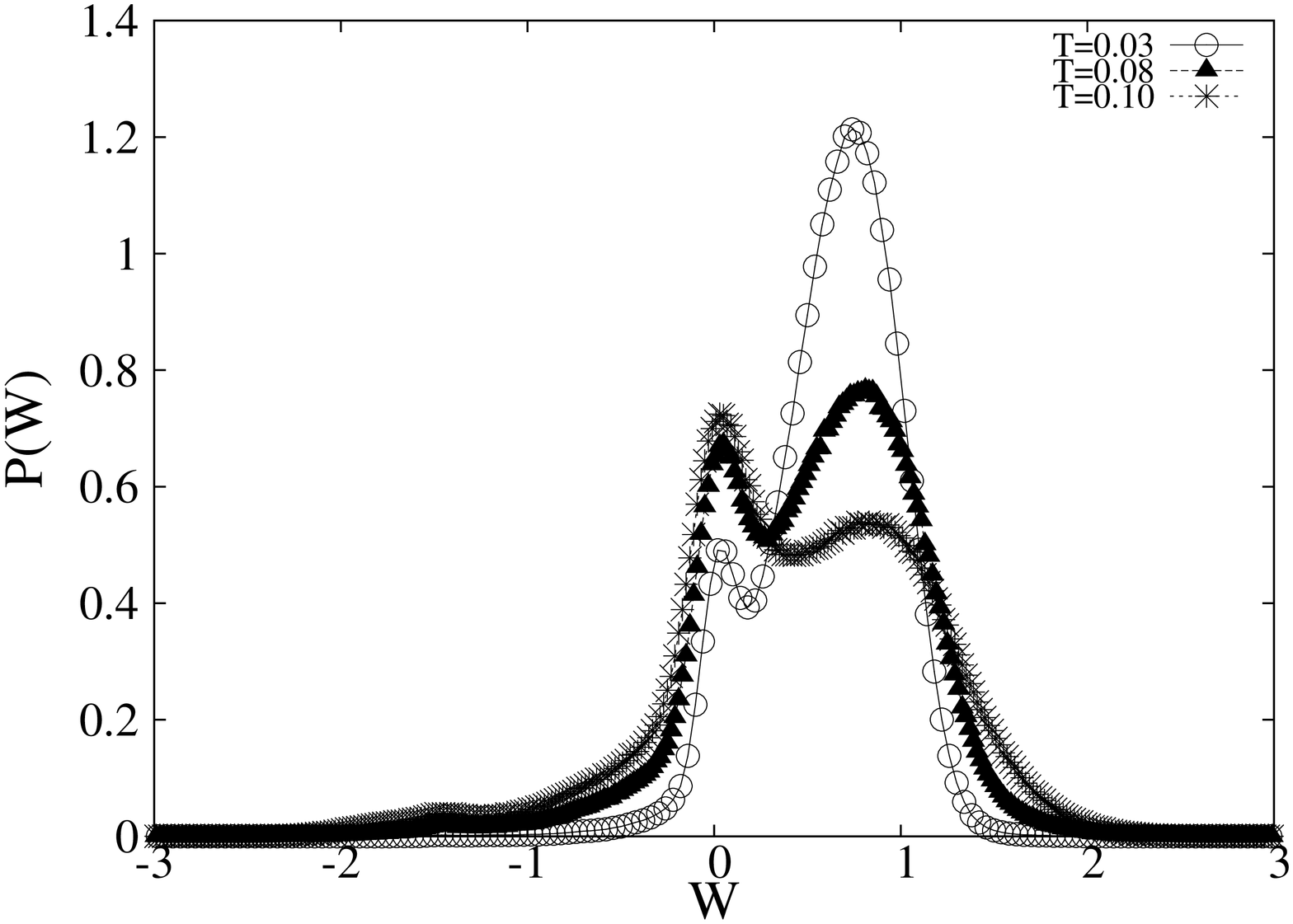}}
\caption{Plot of $P(W_p)$ for different values of $T$ (Fig.a for low $T$ and Fig.b for higher $T$) for the inhomogeneous system;
$\tau=8$, $F_0=0.2$, $\gamma_0=0.12$.}
\label{fig:edge}
\end{figure}

As the temperature is increased to 0.018 the average energy $<\overline{W}>$ 
attains a maximum. At this point the two states become almost equally 
populated, a point in the region of kinetic phase transition\cite{Dykman2}. 
The variation of $<\overline{W}>$ is plotted in Fig.11 as a function of 
temperature. This is a signature of stochastic resonance occurring at 
$T=0.018$. The distribution $P(W)$ again shows the largest asymmetry at the 
same temperature $T=0.018$. Here $P(W)$ is distinctly bimodal in nature, 
Fig.12. For this nonuniform friction case, $P(W)$ behaves in a manner similar
to the case of homogeneous system (Fig.7). However, here the out of phase 
state has the lowest energy unlike in the homogeneous case. At $T=0.018$ we 
have the largest contributing second peak of $P(W)$ to $<\overline{W}>$ at any 
nonzero temperature. These intra-well transitions responsible for SR are 
effectively supported by numerous inter-well transitions in the periodic 
potential system.

\subsubsection{The Hysteresis loop: Area and the phase} 
Since the system is asymmetric and the amplitude $x_0$ are large, at low 
temperatures the hysteresis loops are not perfectly elliptical. Yet, it is 
possible to roughly calculate the average amplitude $x_0$ and phase lags 
$-\overline{\phi}$ of the system response at all temperatures. In Fig.11, 
$x_0$ and $-\overline{\phi}$ are also plotted along with $<\overline{W}>$. It 
is clear from the figure that $<\overline{W}>$, $x_0$, and 
$-\overline{\phi}$ all peak almost at the same temperature. Interestingly, the 
peak value of $-\overline{\phi}$ is about $0.82\pi$. For this system 
$-\overline{\phi}$ satisfies peaking criterion of SR stated using the linear 
response theory\cite{Dykman1}. However, magnitude wise the phase lag 
$-\overline{\phi}$ is off by about $\pi/2$.
\section{Discussion and Conclusion}
A periodic potential system driven by a periodic applied field of small
sub-threshold amplitude at a high frequency, close to the natural frequency of
the periodic potential well bottoms, shows stochastic resonance. Here the 
average input energy per period of the field is considered as the quantifier
of SR. The same quantifier had served SR correctly in bistable systems. 
Moreover, the probability distribution of the input energy exhibits similar 
qualitative behaviour at SR as it shows in the bistable systems. The linear 
response theory calculated frequency dependent mobility was found to show 
exactly similar behaviour as the input energy and was previously termed as a
mere dynamical resonance. It was argued that the period of the drive was too 
small compared to the Kramers time of inter-well static potential barrier 
crossing, thus disqualifying the resonance behaviour from being a genuine 
stochastic resonance. However, we find that such an argument just does not 
hold because, in the dynamical situation, the inter-well transitions become 
quite numerous at the temperature where input energy peaks. Moreover, and 
most importantly, the intra-well trajectories show bistability and the 
obtained resonance is a result of transition between these dynamical states 
effectively helped by inter-well transitions.

There has been two rivalling SR criteria involving the phase lag between the
response and the applied field: one\cite{LGamma} stating that the phase lag 
shows inflection at SR with the phase lag equaling $\pi/4$ and the 
other\cite{Dykman1} that phase lag shows a peak at SR. We find that whereas 
the former criterion is only approximately satisfied in the uniform friction 
case, the latter is satisfied for the nonuniform friction case. Thus, these 
criteria seem to be true for specific systems and hence the phase lag 
$\overline{\phi}$ cannot be taken as a universal quantifier such as the input 
energy for SR.

MCM and AMJ acknowledge partial financial support from BRNS, DAE, India under
Project No. 2009/37/17/BRNS/1959. AMJ thanks DST, Govt. of India for financial
support.

\end{document}